\pdfoutput=1
\documentclass[12pt]{article}
\usepackage{graphics}
\usepackage{graphicx}
\usepackage[active]{srcltx}
\usepackage{float}
\restylefloat{figure}

\begin{document}
\large
\date{ }
\begin{center}
{\Large UCN transport simulation in solid deuterium crystals}

\vskip 0.5cm

Yu.N. Pokotilovski\footnote{e-mail: pokot@nf.jinr.ru}

\vskip 0.5cm
            Joint Institute for Nuclear Research\\
              141980 Dubna, Moscow region, Russia\\

\vskip 0.5cm

{\bf Abstract\\}

\end{center}

\begin{minipage}{130mm}

\vskip 0.5cm

 The extraction efficiency of ultracold neutrons from cryogenic moderators
depends critically on the neutron transparency of the moderator material.
 A Monte Carlo simulation of the probability of the UCN going out from
non-ideal (disordered) solid deuterium crystals has been performed.
 It was based on the use of a correlation function describing the density
fluctuations in a disordered material, the latter being inferred from the
measured very low neutron energy total cross sections for this material.
\end{minipage}

\vskip 0.8cm

 Ultracold neutrons \cite{ucn} (UCN, the energy below $\sim$ 0.3$\mu$eV), which
can be contained in closed volumes, have found important applications in the
investigation of fundamental properties of the neutron \cite{proc}.

 The best UCN source -- ILL UCN turbine \cite{turb} yields maximum UCN density
in a trap of $\sim$ 50 n/cm$^{3}$ and, respectively, the flux
$\sim 5\times 10^{3}$n cm$^{-2}$s$^{-1}$.

 Recent years brought hope for significant progress in the intensity of very
low energy neutron beams.
 This is connected with the possible use of the most effective cold moderators:
solid deuterium or solid deuterocarbons at low temperatures of $\sim$ 5 -- 10 K
for production of very cold neutrons.

 The most promising and popular material for the UCN converter is solid
deuterium.
 The first test of the UCN production in solid deuterium has been performed by
the PNPI group \cite{Ser-80}, which then has been studying experimentally the
UCN production in deuterium during many years [5-11].

 The first estimate of the UCN production and up-scattering cross sections in
solid deuterium was made in \cite{D2-Gol}.
 The phonon frequency spectrum of solid deuterium was calculated in
\cite{Gol-D} using the dispersion relations measured in \cite{disp}.
 Later calculations in the incoherent approximation \cite{incoh} of the UCN
scattering on phonons in solid deuterium reproduced the results of \cite{Gol-D},
this publication also contains the calculation of the UCN spin-flip
up-scattering on para-deuterium with para-ortho transitions of deuterium
molecules.
 The incoherent approximation and the density of states of \cite{Gol-D} was
used in the calculations of the UCN production in \cite{my-calc}.

 It was proposed that because of comparatively short neutron lifetime in
solid deuterium (140 ms in pure ortho-deuterium at T=0 K) solid deuterium can
find better application in a pulsed mode of the UCN production [17,9].

 A solid deuterium UCN source was also demonstrated in the experiments in LANL
\cite{LA}, it is under construction at FRM-2 \cite{FRM} and
is in the stage of completion in PSI \cite{PSI}.

 The incoherent model of neutron interaction in solid deuterium was generally
confirmed in the later works \cite{LA,my-PSI,my-Mainz,Gra,Gut}.

 Recently a re-evaluation of the neutron scattering in solid deuterium was
performed in \cite{coh}, it was based on the calculations in the frame of a
rigorous model of coherent neutron scattering by deuterium nuclei.
 The resulting scattering cross sections obtained in this work are
significantly different from the previous predictions \cite{D2-Gol,Gol-D,incoh}
based on the incoherent approximation.
 Especially large difference between the incoherent approach and coherent one
was found in the case of UCN up-scattering with phonon annihilation in the
temperature range of 5-10 K: the first one 3-4 times surpassed the second one.
 These new calculations found some experimental confirmation in \cite{Lav}.

 It was clear from the very beginning that structural homogeneity of a
deuterium crystal is very important for the UCN yield from the UCN converter.
 The neutron scattering on structural inhomogeneities increases the time
neutrons spend inside the UCN converter before leaving it, increasing the
probability of their capture and up-scattering.
 It was discovered in \cite{Ser-cr} that the transmission of
very low energy neutrons through solid deuterium depends critically on the way
of preparation of a solid deuterium sample.
 The light transmission investigations of solid deuterium crystals were
performed in \cite{PSI-light}.
 It was demonstrated that optical quality of crystals depends on the speed
of crystal growing - more transparent crystals were obtained when they were
carefully grown over a time period of about 12 hours at temperature close to
the triple point.

 The total cross section in the UCN energy range for solid deuterium
frozen from liquid state was measured in \cite{sD-cross}.
 These measurements showed that the best transmission have the solid deuterium
samples prepared in the process of very slow cooling - less than 1 K per hour.
 This procedure of obtaining good quality deuterium crystals was based on
preliminary investigations described in \cite{PSI-light}.
 Temperature cycling between 5 K and 10 K, and between 5 K and 18 K seriously
deteriorated the neutron transmission.
 It is obvious that crystal imperfections introduced by thermal cycling
increased the UCN scattering.
 In contrast to this experience, the solid deuterium crystals used as UCN
converters and condensed directly from gas phase significantly increased the
UCN yield being subjected to repeated thermal cycling \cite{my-Mainz}.

 We consider solid deuterium crystals frozen in the process of very slow
cooling in \cite{sD-cross} as the most perfect of all possible samples prepared
from liquid, and use the energy dependence of the total cross section
(Fig. 2 of \cite{sD-cross}) to infer the parameters characterizing structural
inhomogeneity of the sample.
 These parameters are used further to simulate the probability for the UCN
generated in a similar solid deuterium UCN source, to go out from the source to
vacuum.

 To obtain the UCN scattering cross section on structural inhomogeneities
the experimental points of Fig. 2 of Ref. \cite{sD-cross} were corrected for the incoherent elastic scattering
on deuterium (2.05 $b$), the UCN capture by deuterium (1.19 $b/v$) and hydrogen (733 $b/v$) nuclei, the UCN
up-scattering in the result of one-phonon annihilation (0.83 $b/v$), and the spin-flip up-scattering on
para-deuterium (123 $b/v$), where $v$ is the neutron velocity in medium in m/s.
 The cross sections of the latter processes were taken from \cite{incoh}.
 All the microscopic cross sections here and further are given per one atom of respective element.
 The concentrations of hydrogen (0.05\%) and para-fraction (1.4\%) were taken as
they were done in \cite{sD-cross}.
 The atomic density of solid deuterium at 5 K was taken to be
$6\times 10^{22}$ cm$^{-3}$.
 The resulting neutron wavelength dependence of the UCN scattering on
inhomogeneities is shown in Fig. 1.

 The transmission of very slow neutrons through an inhomogeneous medium was
studied previously by A. Steyerl \cite{Ste}.
 It was demonstrated that transmission as a function of neutron wave length may
be used to deduce characteristic parameters of inhomogeneities, in particular
their size and density.

 It is known that in the Born approximation the differential macroscopic cross
section of elastic scattering for neutrons transmitting through an isotropic
inhomogeneous medium has the form \cite{Step}:
\begin{equation}
\frac{d\Sigma_{el}}{d\Omega}=\frac{1}{\pi}\Biggl(\frac{m}{\hbar^{2}}\Biggr)^{2}
\int_{0}^{\infty}G(\rho)\frac{sin(q\rho)}{q\rho}\rho^{2}d\rho,
\end{equation}
where m is the neutron mass, q=$|\vec k'-\vec k|$ is the neutron wave vector
change, and G($\vec r, \vec r'$)=$<\delta$U($\vec r$) $\delta$U($\vec r')>$,
($\rho=|\vec r'-\vec r|$) is the correlation function of fluctuations of the
local neutron-medium interaction potential.

The latter is
\begin{equation}
U=\frac{\hbar^{2}}{2m}\sum_{i} 4\pi N_{i}b_{i},
\end{equation}
where N$_{i}$ is the atomic density and b$_{i}$ is the coherent scattering
lengths of nuclei of the medium, so that
$\delta$U($\vec r$)=U($\vec r$) - $<$U($\vec r)>$.

The total cross section after integration over solid angle with the account of
the solid angle of the neutron detector is:
\begin{equation}
\Sigma_{el}(k)=2\Biggl(\frac{m}{\hbar^{2}}\Biggr)^{2}\frac{1}{k^{2}}
\int_{0}^{\infty}G(\rho)[cos(2k\rho\cdot sin\theta_{0})- cos(2k\rho)]d\rho,
\end{equation}
where 2$\theta_{0}$ is the angle of the neutron detector from the sample.

For the exponential correlation function
\begin{equation}
G(\rho)=G_{0}e^{-\rho/\rho_{0}},
\end{equation}
where $\rho_{0}$ is the correlation length, we have the expression
\begin{equation}
\Sigma_{el}(k)=2\Biggl(\frac{m}{\hbar^{2}}\Biggr)^{2}\frac{G_{0}\rho_{0}}{k^{2}}
\Bigl[\frac{1}{1+4k^{2}\rho_{0}^{2}sin^{2}\theta_{0}}- \frac{1}{1+4k^{2}\rho_{0}^{2}}\Bigr]
\end{equation}
for the total cross section, and
\begin{equation}
\frac{d\Sigma_{el}}{dq}=\frac{4}{k^{2}}\Biggl(\frac{m}{\hbar^{2}}\Biggr)^{2}
\frac{G_{0}q\rho_{0}^{3}}{1+q^{2}\rho_{0}^{2}}
\end{equation}
for the differential cross section.

 This formalism was used for the interpretation of the measurements of the
total cross sections for solid deuterium at 5 K \cite{sD-cross}.
 The parameters of the correlation function of Eqs. (4) and (5) were inferred
by the least squared method from the corrected experimental points for the
macroscopic cross section of the UCN scattering on inhomogeneities of Fig. 1.
 The inferred parameters of the exponential correlation function were found
to be: $G_{0}=(0.33\pm 0.08)$ neV$^{2}$, $\rho_{0}=(28\pm 3.3)$\,\AA.
 The total cross section of Eq. (5) with these parameters is shown in Fig. 1.

 Fig. 2 shows the macroscopic total and transport cross sections of scattering
on inhomogeneities:
\begin{equation}
\Sigma_{tr}=\int (1-cos{\theta})d\Sigma,
\end{equation}
where $\theta$ is the neutron scattering angle.

 The difference between total and transport cross sections is about 5\% at the
neutron energy of 200 neV and is lower at lower energies, it follows that the
scattering by inhomogeneities is almost isotropic in this energy range.

 The differential cross section of Eq. (6) with inferred parameters $G_{0}$ and
$\rho_{0}$ was used to perform Monte Carlo simulation of the UCN transport in
a cylindrical deuterium crystal aiming to obtain the probability for the UCN
to go out of the crystal.

 The Monte Carlo program used here is the development of our program used
earlier for simulation of transport of very slow neutrons in matter, in
particular in cold moderators with their temperature changing during UCN motion
in a moderator \cite{gran}.

 The UCN were assumed to be born homogeneously with isotropic angular
distribution, the reflection from the side walls of a cylinder was specular,
from the top and bottom walls -- according to the cosine law.
 The quantum reflection from the boundary deuterium-vacuum at the exit side of
the sample was taken into account according to:
\begin{equation}
R=\Biggl(\frac{v_{0,\perp}-v_{\perp}}{v_{0,\perp}+v_{\perp}}\Biggr)^{2},
\end{equation}
where $R$ is the neutron reflection probability, $v_{\perp}$ is the normal to
the boundary component of the neutron velocity in medium,
$v_{0,\perp}=(v_{\perp}^{2}+v_{b}^{2})^{1/2}$ is the normal to the boundary
component of the neutron velocity in vacuum, $v_{b}$=4.46 m/s is the calculated boundary
velocity of deuterium, the boundary energy $E_{b}=104$\,neV.
 Recent measurement \cite{Mainz-b} of the minimal neutron energy from the
solid deuterium source $E_{min}=(99\pm 7)$\,neV, which should correspond to the
boundary energy of deuterium, does not contradict to this value.

 Two variants for the thickness of the solid deuterium sample were chosen:
15 cm - mean thickness of the solid deuterium source in PSI \cite{PSI} and 8 cm
- typical for the UCN source at the TRIGA reactor of Mainz University
\cite{my-calc,my-Mainz}.

 The UCN up-scattering due to phonon annihilation was calculated in two
variants: the incoherent model \cite{incoh} and coherent one \cite{coh}.
 The result of this simulation as the probability for the UCN to go out from
the cylindrical solid deuterium crystal are shown in Figs. (3-8).

 They illustrate effects of temperature, concentration of para-fraction
in deuterium, and inhomogeneity of solid deuterium on the probability for the
neutrons to leave the moderator.

 Fig. 3 shows the probability for the UCN to go out from perfectly homogeneous
solid deuterium cylinder with a thickness of 15 cm at different temperatures, at the
concentration of para-fraction $c_{p}=0$, the up-scattering on phonons was
calculated according to the incoherent model of Ref. \cite{incoh} and the
coherent model of Ref. \cite{coh}.
 The following one-phonon annihilation cross sections were used: for the
incoherent variant 0.83 b/v, 2.78 b/v, and 7.9 b/v for temperatures 5, 7, and
9 K, respectively; for the coherent variant 0.2 b/v, 0.87 b/v, and 2.78 b/v
for the same temperatures.
 Further the effect of phonon upscattering on the UCN yield will be discussed
only for the variant of calculations in the incoherent model, the calculations according to
the coherent model predict much lower effect of the phonon annihilation
on the UCN loss in a moderator.
 It is seen, that the phonon upscattering effect of heating of an ideal deuterium
crystal from 5 to 9 K decreases the UCN yield more than 3 times down to
$\sim$6.5\% of produced neutrons with an energy 150 neV.

 Fig. 4 shows the same as in Fig. 3 (thickness of 15 cm, $c_{p}=0$), but for the
inhomogeneous solid deuterium with the inferred parameters of the exponential
correlation function $G_{0}=0.33$\,neV$^{2}$, $\rho_{0}=28$\,\AA.
 The effect of inhomogeneities is not very essential for this rather good
crystal: neutron yield is decreased 1.3 times and 1.2 times at temperatures 5 and 9 K,
respectively, compared to an ideal crystal for the same UCN energy.

 The effect of spin-flip upscattering due to admixture of para-deuterium in
an inhomogeneous crystal with a thickness of 15 cm at 9 K is seen in Fig. 5.
 Compared to the effect of phonon annihilation, additional influence of the UCN
upscattering with para-ortho transition of deuterium molecules is not
as dangerous: at 2\% of para-deuterium
the UCN yield decreases 22\% for neutrons at 150 neV.

 Naturally, all the neutron loss effects are larger for thicker sample.
 Figures 6, 7, and 8 show the neutron energy dependence of the effects of
temperature, concentration of para-fraction in deuterium, and inhomogeneity of
the crystal on the probability for the neutrons to leave the deuterium
moderator with a thickness of 8 cm.

 Fig. 6 shows the probability for the UCN to go out from a perfectly homogeneous
solid deuterium cylinder with a thickness of 8 cm at different temperatures, at the
concentration of para-fraction $c_{p}=0$.
 In this case heating of an ideal deuterium crystal from 5 to 9 K decreases the
UCN yield due to the phonon-upscattering effect the same 3 times, but the
absolute values of the probability for the neutrons at 150 neV to
leave the moderator are larger: 0.36 and 0.12 compared to 0.21 and 0.065 at
these temperatures at a thickness of 15 cm.

 Fig. 7 shows the same as in Fig. 4 (inhomogeneous solid deuterium, $c_{p}=0$),
but for a thickness of 8 cm.
 The effect of inhomogeneities is not essential in this case too: neutron
yield is decreased 1.24 times and 1.2 times at temperatures 5 and 9 K,
respectively, compared to an ideal crystal for the same UCN energy.

 And, at last, Fig. 8 shows the effect of spin-flip upscattering due to
admixture of para-deuterium in inhomogeneous crystal with a thickness of 8 cm at
9 K.
 Compared to the phonon annihilation additional UCN upscattering with
para-ortho transition of the deuterium molecules is not as dangerous: at a concentration of para-fraction 2\%
the UCN yield decreases approximately the same 20\% for neutrons at 150 neV.

 Maximum UCN losses are seen if we compare, for example, the probability for
the neutrons to leave the sample at the ideal condition of homogeneous material at T=0 and in the worst shown
variant: inhomogeneous solid deuterium at T=9 K, concentration of para-fraction $c_{p}=0.02$.
 In this case for the UCN at 150 neV the yield is 6.5 times
lower than from the ideal crystal with a thickness of 15 cm and is only 4.5\% of
produced neutrons.
 At temperature of 5 K the UCN yield from a homogeneous crystal is 4.2-4.7
times larger than from a inhomogeneous one at 9 K depending on the
para-fraction concentration.
 For deuterium thickness 8 cm maximum overall losses are lower than for 15 cm:
the yield is about 8\% of produced neutrons at 9 K and 2\% concentration of para-deuterium molecules.

 The author is grateful to Dr. Malgorzata Kasprzak for sending the data tables
for Fig. 2 of Ref. \cite{sD-cross}.

\newpage

\begin{figure}
\begin{center}
\resizebox{13cm}{13cm}{\includegraphics[width=\columnwidth]{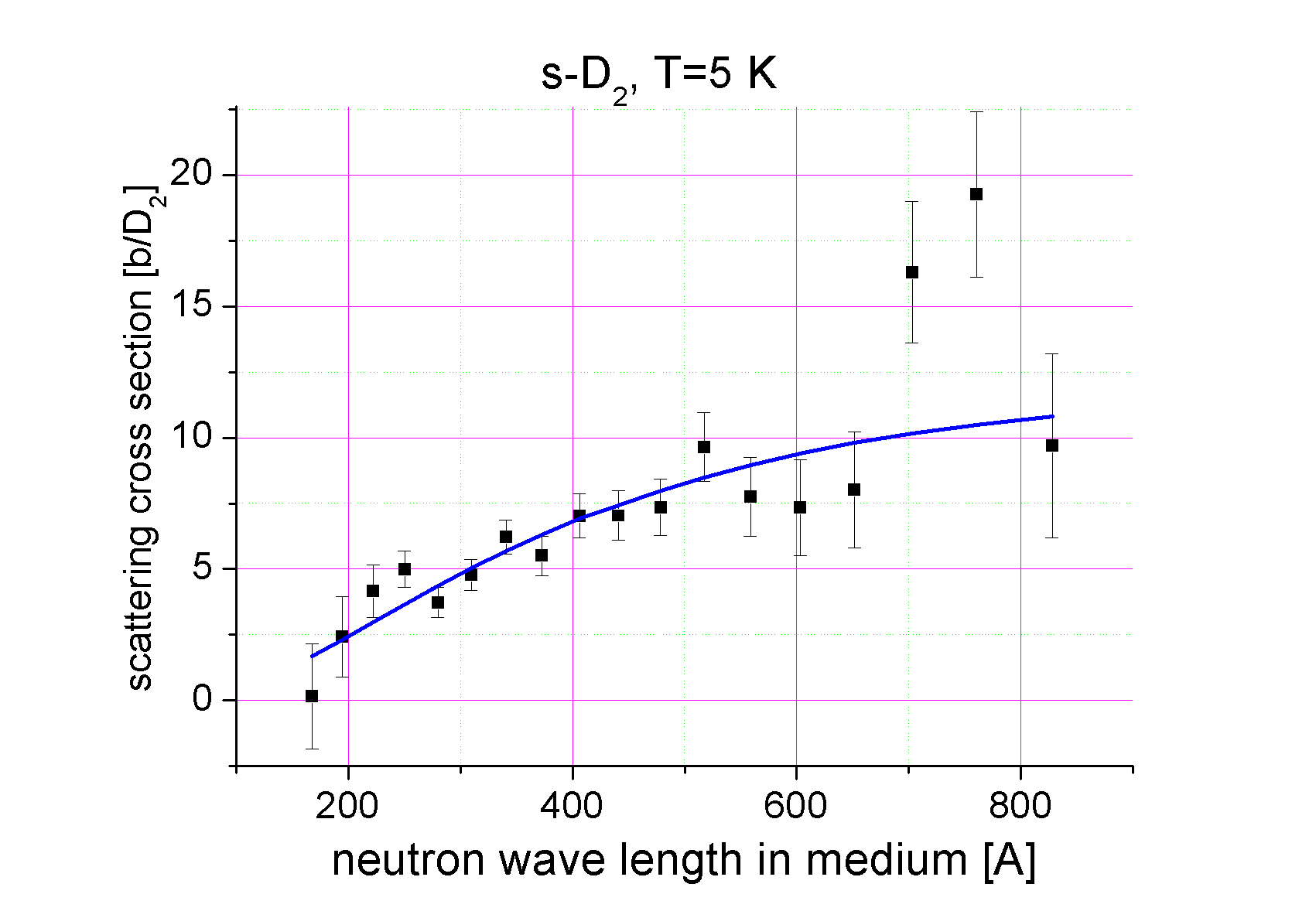}}
\end{center}
\caption{Cross section of scattering on inhomogeneities for the deuterium
sample at 5 K (Fig. 2 of \cite{sD-cross}) as a function of the in-medium
neutron wave length after subtraction of the neutron capture, up-scattering,
and the elastic incoherent scattering by deuterium nuclei.
 The curve is the approximation according to Eq. (5) with the deduced
parameters $G_{0}$=0.33 neV$^{2}$, $\rho_{0}$=28\,\AA.}
\end{figure}

\begin{figure}
\begin{center}
\resizebox{13cm}{13cm}{\includegraphics[width=\columnwidth]{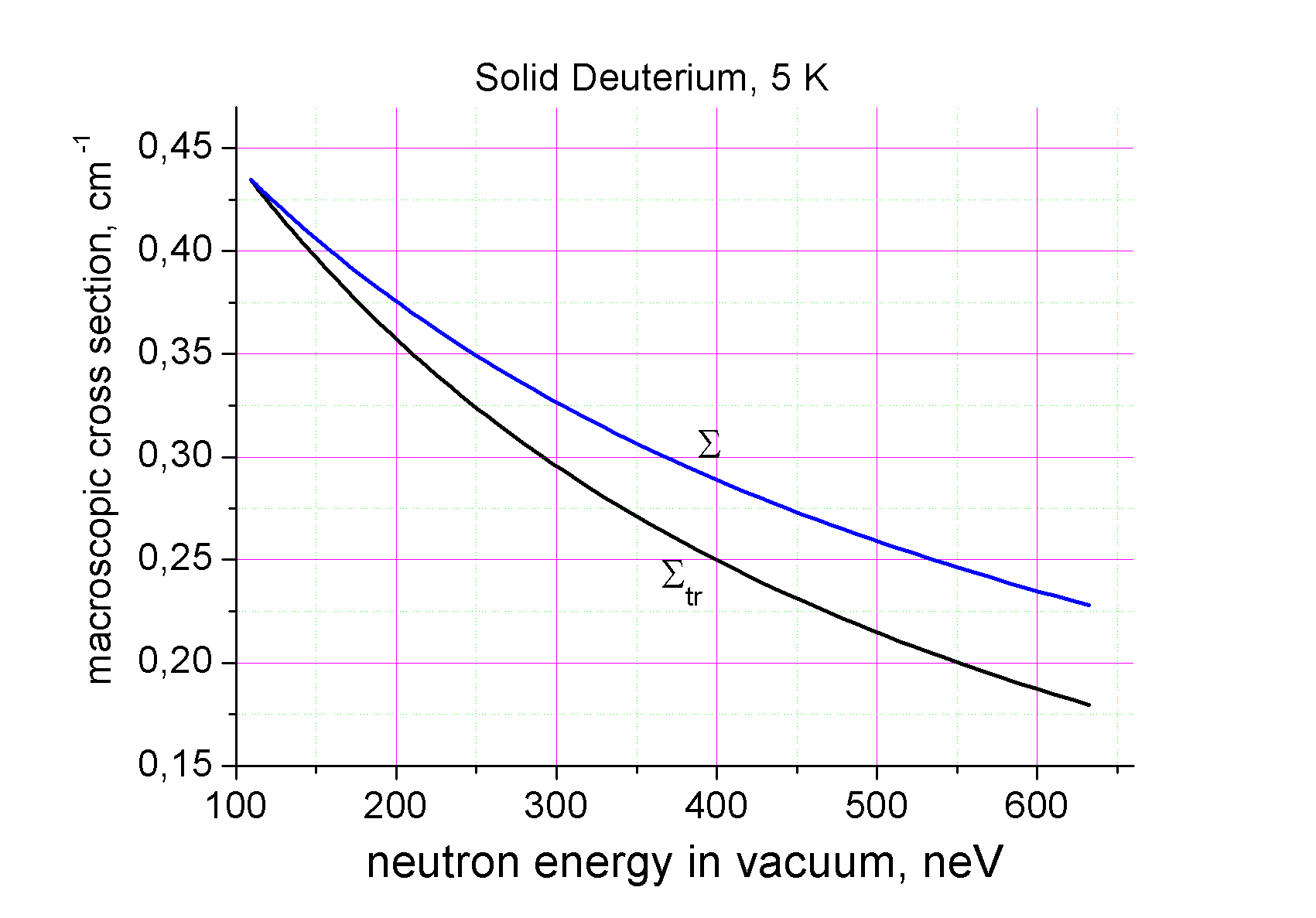}}
\end{center}
\caption{Macroscopic total and transport cross sections of neutron scattering by the solid deuterium crystal
inhomogeneities according to Eqs. (5) and (6) with the deduced parameters $G_{0}$=0.33 neV$^{2}$,
$\rho_{0}$=28\,\AA.}
\end{figure}

\begin{figure}
\begin{center}
\resizebox{13cm}{13cm}{\includegraphics[width=\columnwidth]{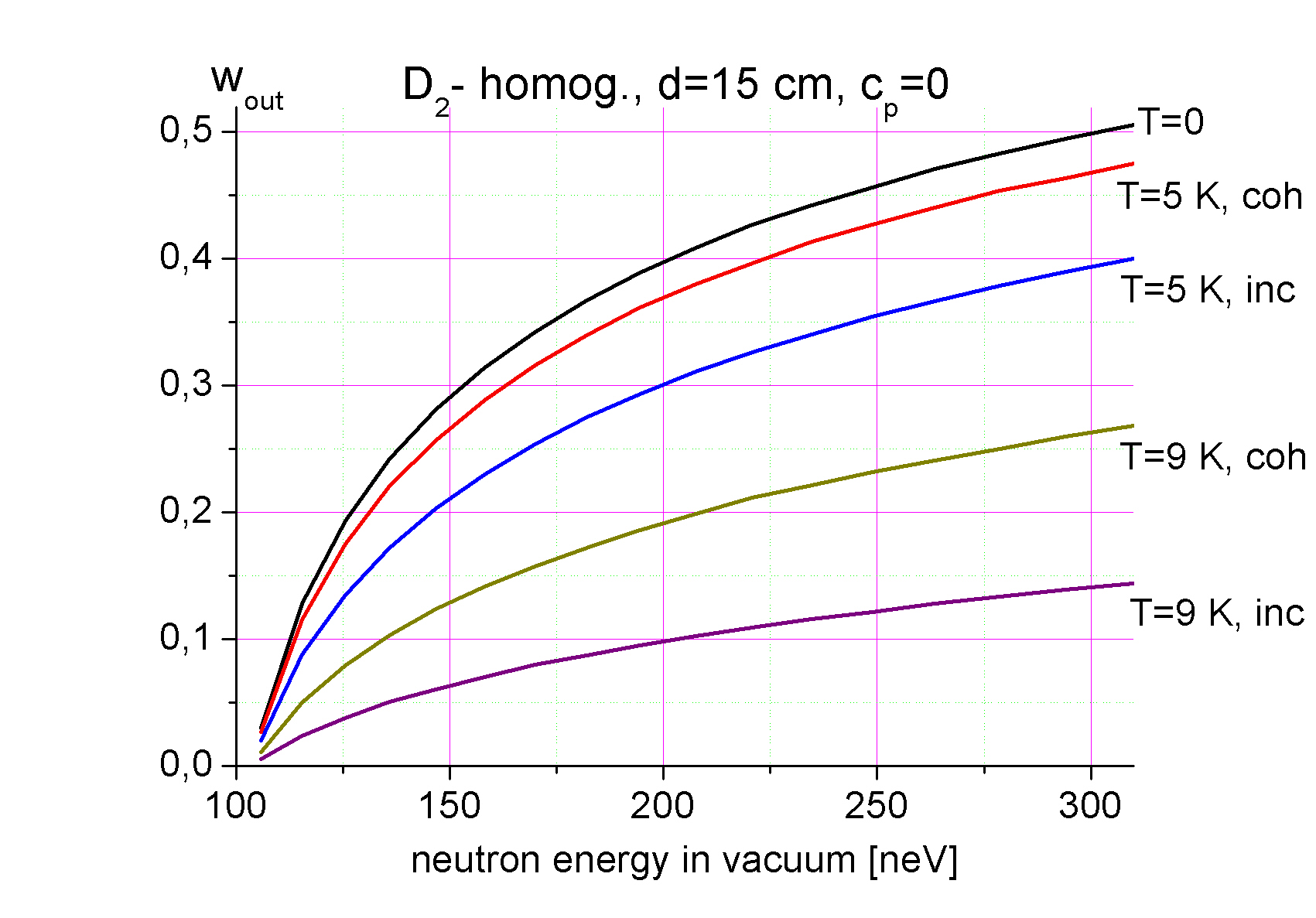}}
\end{center}
\caption{The probability for the UCN to go out from the homogeneous solid
deuterium cylinder with thickness 15 cm, the concentration of para-fraction
$c_{p}=0$, the up-scattering on phonons was calculated according to the
coherent model of Ref. \cite{coh} and the incoherent model of Ref. \cite{incoh}.}
\end{figure}

\begin{figure}
\begin{center}
\resizebox{13cm}{13cm}{\includegraphics[width=\columnwidth]{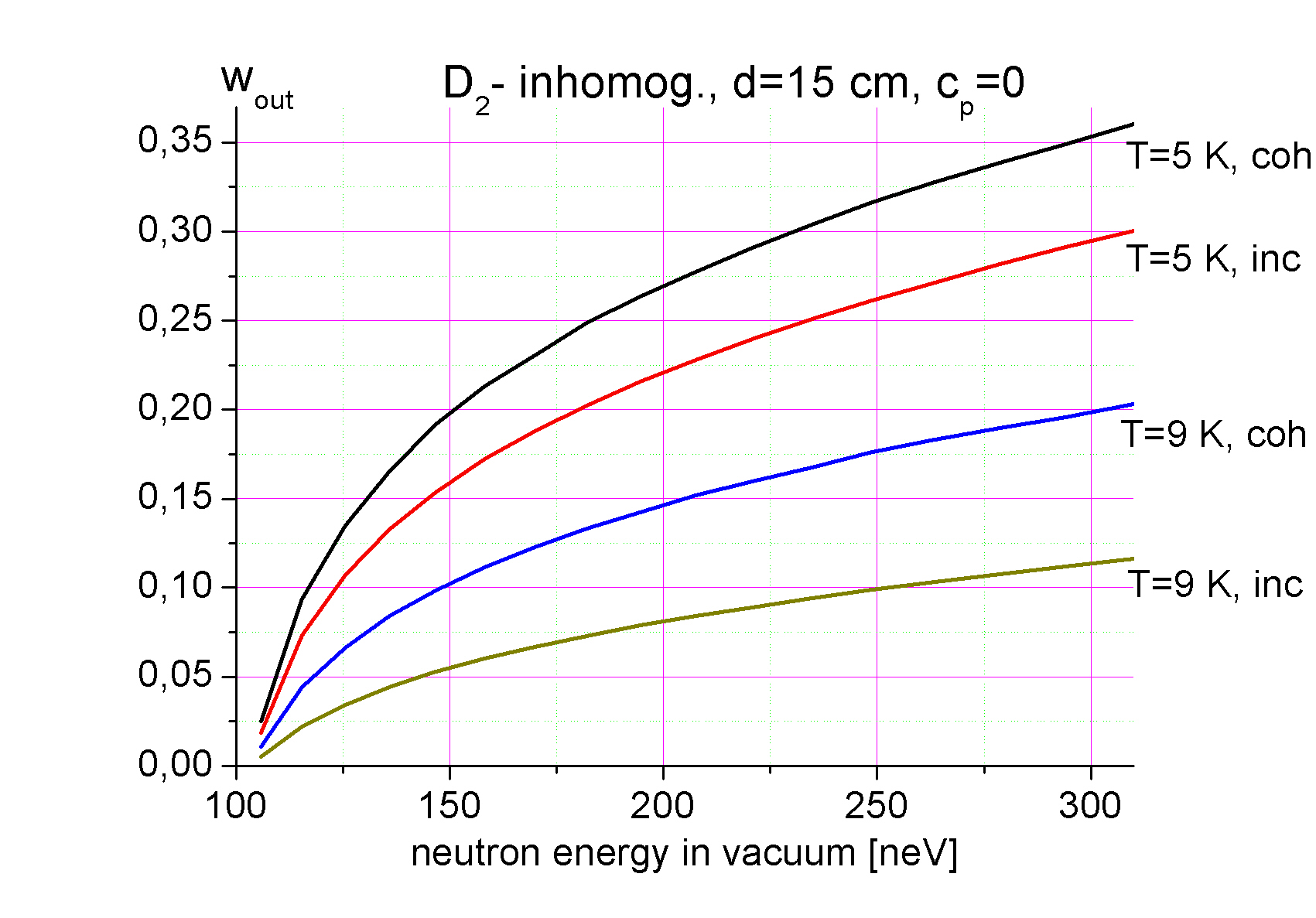}}
\end{center}
\caption{The same as in Fig. 3, but for the inhomogeneous solid deuterium with
the inferred parameters of exponential correlation function
$G_{0}=0.33$\,neV$^{2}$, $\rho_{0}=28$\,\AA}.
\end{figure}

\begin{figure}
\begin{center}
\resizebox{13cm}{13cm}{\includegraphics[width=\columnwidth]{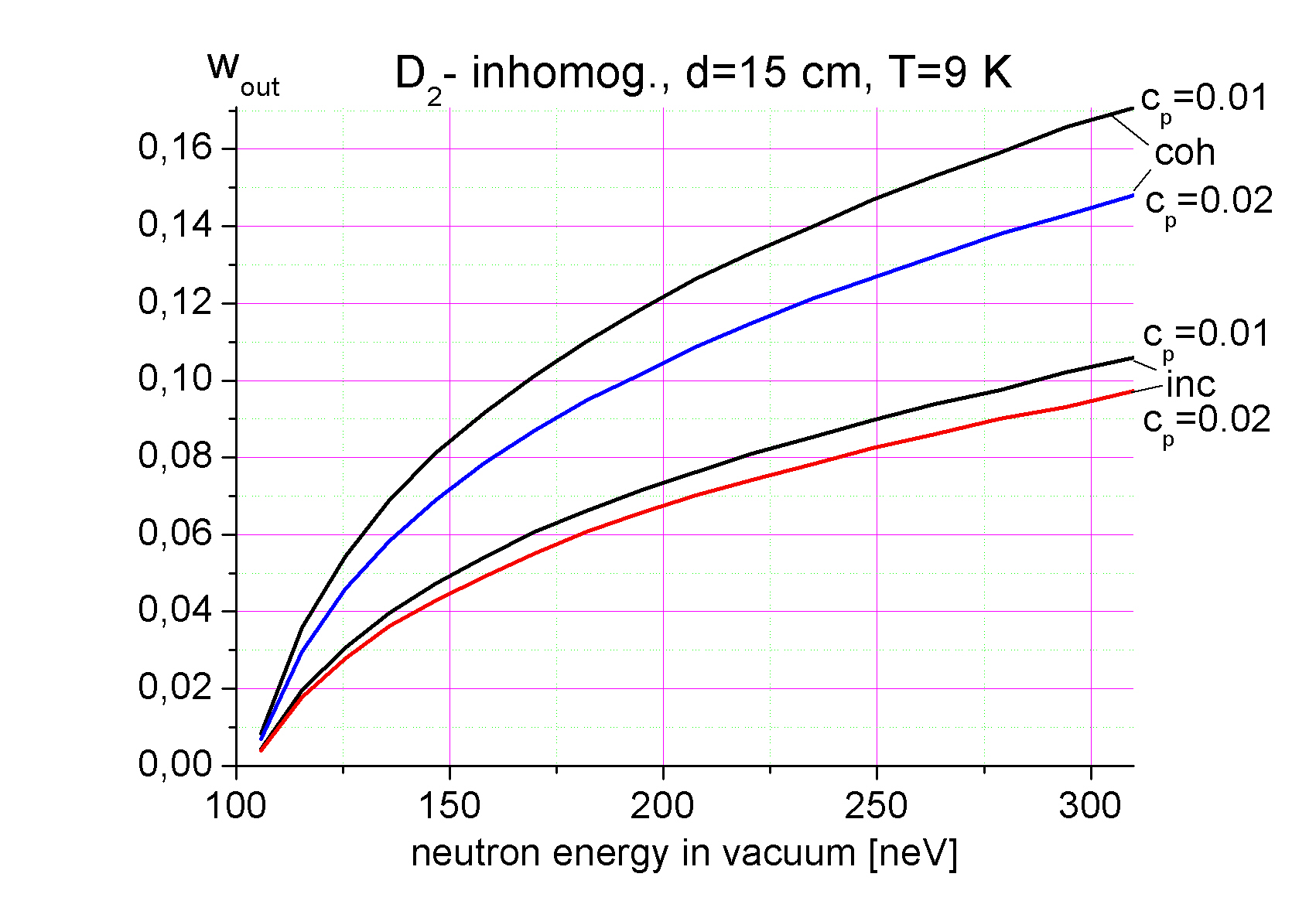}}
\end{center}
\caption{The probability for the UCN to go out from the inhomogeneous solid
deuterium crystal with thickness 15 cm and temperature T=9 K, at two
concentrations of para-fraction $c_{p}=0.01$ and 0.02, the up-scattering on
phonons was calculated according to the coherent model of Ref. \cite{coh} and
the incoherent model of Ref. \cite{incoh}.}
\end{figure}

\begin{figure}
\begin{center}
\resizebox{13cm}{13cm}{\includegraphics[width=\columnwidth]{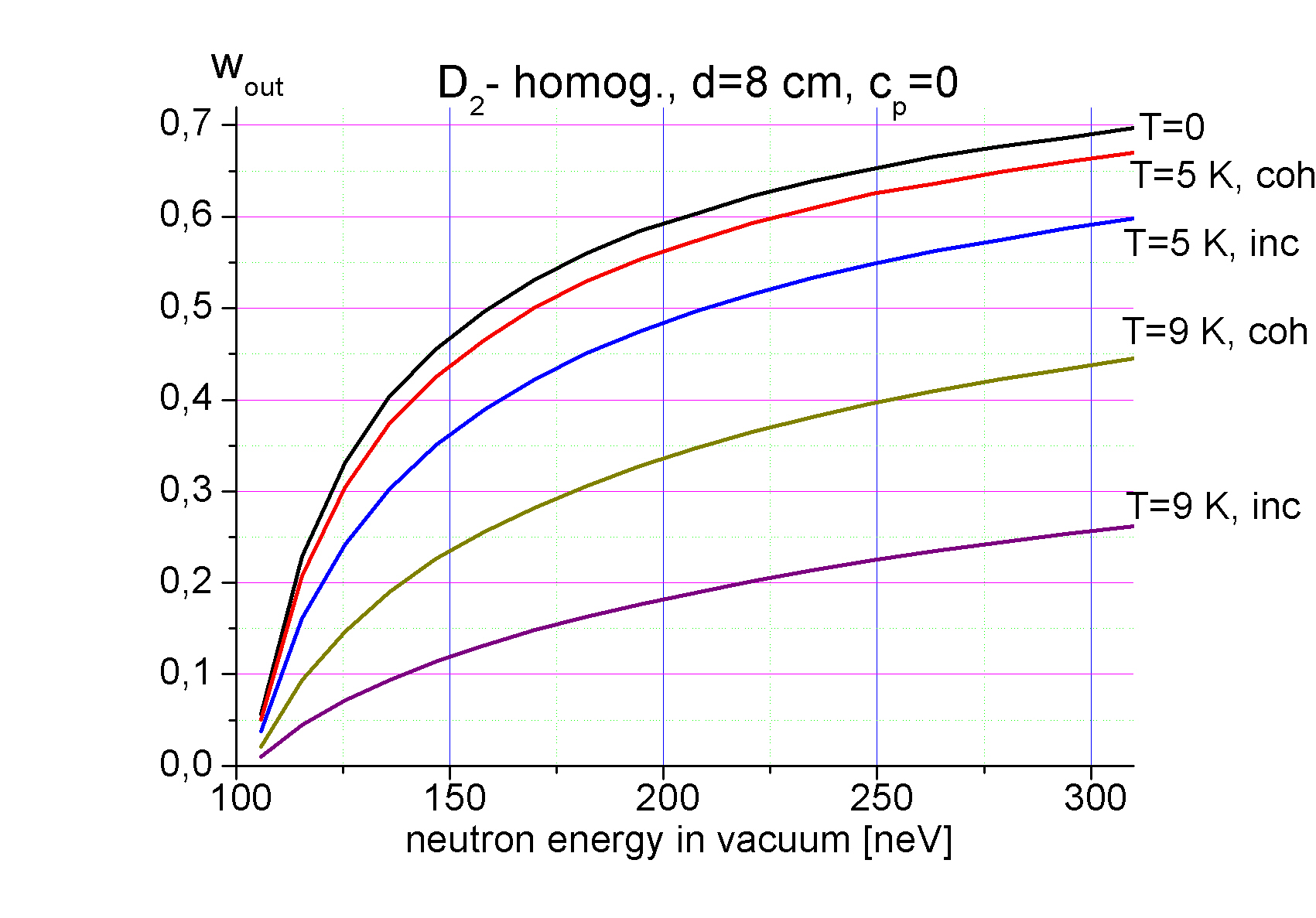}}
\end{center}
\caption{The same as in Fig. 3, but for the deuterium thickness 8 cm.}
\end{figure}

\begin{figure}
\begin{center}
\resizebox{13cm}{13cm}{\includegraphics[width=\columnwidth]{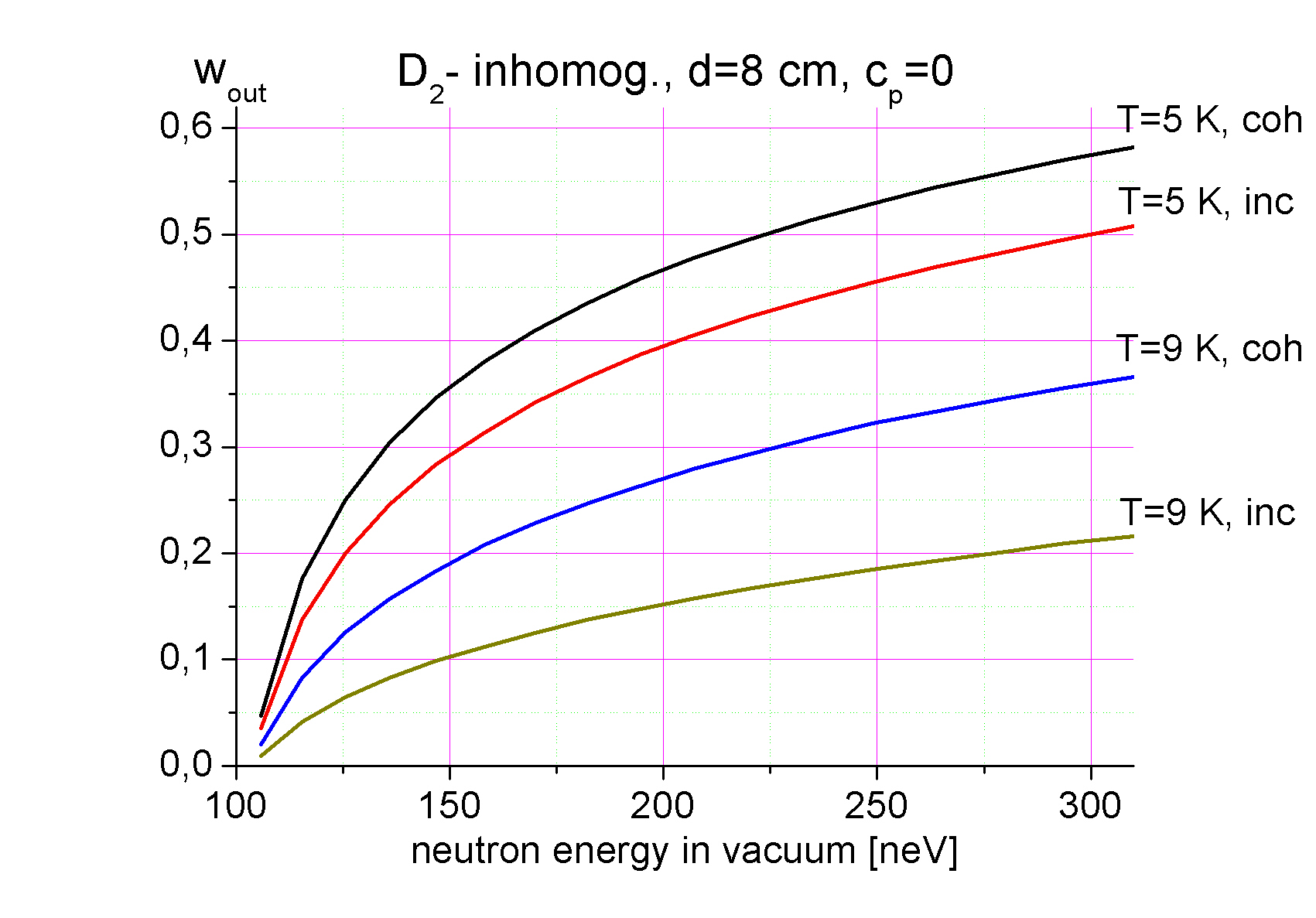}}
\end{center}
\caption{The same as in Fig. 4, but for the deuterium thickness 8 cm.}
\end{figure}

\begin{figure}
\begin{center}
\resizebox{13cm}{13cm}{\includegraphics[width=\columnwidth]{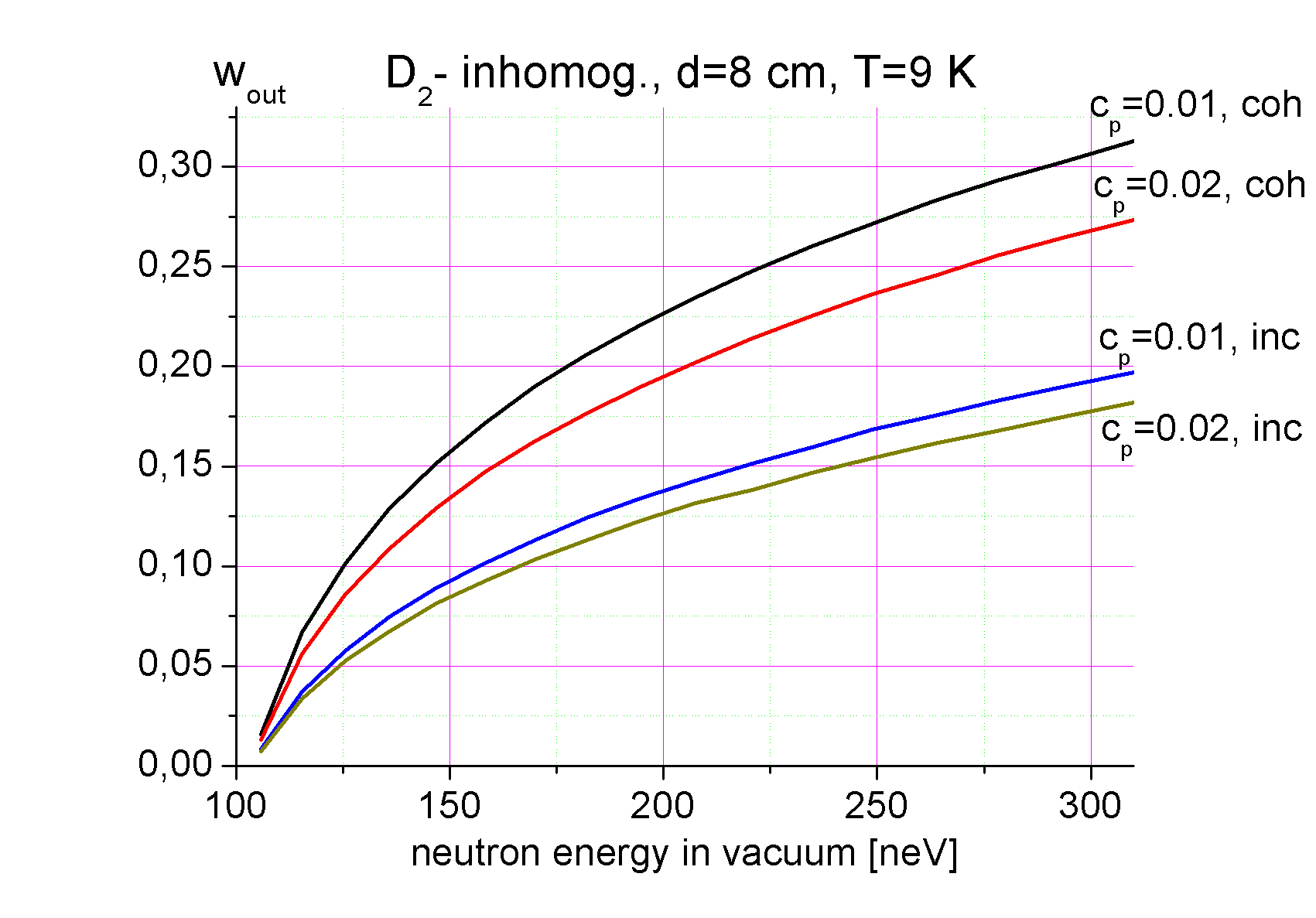}}
\end{center}
\caption{The same as in Fig. 5, but for the deuterium thickness 8 cm.}
\end{figure}

\end{document}